\documentclass[12pt]{extarticle}
\usepackage{times}
\usepackage{graphicx}
\usepackage{array, ltablex, multirow}
\usepackage[numbers,square,sort&compress]{natbib}
\usepackage{multirow}
\usepackage{float}
\usepackage{authblk}
\usepackage{xcolor}
\usepackage{caption}

\usepackage{cancel}
\title{Atomic structure of intrinsic and electron-irradiation-induced defects in MoTe$_{2}$}
\author[1]{Kenan Elibol}
\author[1]{Toma Susi}
\author[1]{Giacomo Argentero}
\author[1]{Mohammad Reza Ahmadpour Monazam}
\author[1]{Timothy John Pennycook}
\author[1]{Jannik C. Meyer}
\author[1$^\ast$]{Jani Kotakoski}
\affil[1]{Faculty of Physics, University of Vienna, Boltzmanngasse 5, A-1090, Vienna, Austria}
\affil[$^\ast$]{Corresponding author; E-mail: jani.kotakoski@univie.ac.at}
\date{}
\begin{document}
	\maketitle
	
	\begin{abstract} 
		Studying the atomic structure of intrinsic defects in two-dimensional transition metal di\-chal\-co\-ge\-ni\-des is difficult since they damage quickly under the intense electron irradiation in transmission electron microscopy (TEM). However, this can also lead to insights into the creation of defects and their atom-scale dynamics. We first show that MoTe$_2$ monolayers without protection indeed quickly degrade during scanning TEM (STEM) imaging, and discuss the observed atomic-level dynamics, including a transformation from the 1H phase into 1T$'$, three-fold rotationally symmetric defects, and the migration of line defects between two 1H grains with a 60$^\circ$ misorientation. We then analyze the atomic structure of MoTe$_2$ encapsulated between two graphene sheets to mitigate damage, finding the as-prepared material to contain an unexpectedly large concentration of defects. These include similar point defects (or quantum dots, QDs) as those created in the non-encapsulated material, and two different types of line defects (or quantum wires, QWs) that can be transformed from one to the other under electron irradiation. Our density functional theory simulations indicate that the QDs and QWs embedded in MoTe$_{2}$ introduce new midgap states into the semiconducting material, and may thus be used to control its electronic and optical properties. Finally, the edge of the encapsulated material appears amorphous, possibly due to the pressure caused by the encapsulation.
\end{abstract}
	
	\newpage\section*{Introduction}
	Knowing the atomic arrangement of materials beyond the crystal structure is imperative for understanding their real-life  performance, as defects often have a profound influence on material properties. The recent combination of advances in aberration corrected transmission electron microscopy (TEM) and two-dimensional (2D) materials has provided an unprecedented possibility both to directly image defects and to control them~\cite{Susi2017}. Among 2D materials, the family of transition metal dichalcogenides (TMDC) with the chemical structure MX$_2$ (M is a transition metal and X a chalcogen atom) displays arguably the widest range of material properties. Remarkably, even the same TMDC in its different configurations can exhibit different behaviors. For example, 2H, 1T, 1T$'$, and T$_d$ phases of MoTe$_{2}$ all have different electronic characteristics (semiconductor, metal, semimetal, and superconductor, respectively)~\cite{Deng2016,Han2016,Joshi2016,Diaz2015}. Also the electron mobilities of the different phases vary: 1T$'$ has a mobility of 4000 cm$^{2}$/Vs, whereas that of 2H is two orders of magnitude lower~\cite{Keum2015}. Further, similar to other TMDCs, although monolayer MoTe$_{2}$ is a direct bandgap semiconductor, its bulk form has an indirect bandgap~\cite{Kuiri2016}.
	
	Of the different MoTe$_2$ phases, the normally metastable 1T$'$ becomes stable under strain against the otherwise more stable 2H~\cite{Park2015, Duerloo2014}. At temperatures below 240~K, 1T$'$ is expected to transition into the superconducting T$_d$ phase~\cite{Sun2015}. Until now, different phases have been obtained in MoTe$_2$ through methods including strain and exposure to electron beams, light and different temperatures~\cite{Li2016,Huang2016,Cho2015,Song2016,Zhang2016}. However, the degree of control regarding the location and size of the transformed regions has remained poor.
	
	Defects such as vacancies, grain boundaries, and edges have a significant impact on the physical properties of TMDCs. For example, it is possible to induce n-type or p-type transport in MoS$_{2}$ via sulfur or molybdenum vacancies~\cite{Lin2016}. Looped line defects consisting of 8-5-5-8 membered rings in MoS$_{2}$ and WSe$_{2}$ induce mid-gap states~\cite{Lin2015,Enyashin2013,Ghorbani-Asl2013}, as do grain boundaries consisting of 8-4-4 or 8-4-4-8 membered rings in MoS$_{2}$~\cite{VanderZande2013,Dumcenco2015}. Mirror-symmetric grain boundaries in MoSe$_{2}$ and MoTe$_{2}$, studied both theoretically and experimentally by scanning tunneling spectroscopy, also show localized states within the band gap~\cite{Ma2017,Lin2016,Diaz2016}, in contrast to the mirror-symmetric grain boundary in MoS$_{2}$ which acts as a 1D metallic quantum wire~\cite{Zhou2013}. Similarly, the optical properties of TMDCs are altered by defects because the response is directly related to the electronic band structure~\cite{Wang2012}. The edges of TMDCs also have attractive optical properties since the intensity of visible light emitted by them is as high or even higher than the entire bulk of the crystal~\cite{Lin2016}. Further, dopants in TMDCs induce localized states in the electronic band structure and also shift the photoluminescence energy~\cite{Lin2016}. Recent studies demonstrate that producing atomic vacancies via electron- or ion-beam irradiation can enhance ferromagnetism in MoS$_{2}$~\cite{Lin2016}, and that grain boundaries consisting of 5-7 dislocation cores exhibit ferromagnetic behavior whereas those with 4-8 dislocation cores are antiferromagnetic~\cite{Zhang2013,Lin2016}. Despite such efforts towards finding new physical properties and functionalities in various 2D materials, the atomic structure of defects in MoTe$_{2}$ remains poorly understood.  
	
	In this work, we study the atomic structure of MoTe$_{2}$ samples consisting of mechanically exfoliated mono-, bi- and triple-layer areas with scanning transmission electron microscopy (STEM). Although the as-exfoliated material deteriorates quickly under electron irradiation due to ionization damage, we were able to image local phase transitions between the 1H and 1T$'$ phases, the formation of three-fold rotationally symmetric defects consisting of 4-8-4 membered rings at high electron doses, as well as the migration of grain boundaries between areas with opposite crystalline orientations (1H grains with misorientation of 60$^\circ$). For further studies, we encapsulated MoTe$_2$ between two graphene monolayers to both protect the material from oxidation and to mitigate damage during STEM imaging. We show that the as-prepared samples contain three-fold rotationally symmetric quantum dots (QDs), and both reflection symmetric (QW1) and two-fold rotational symmetric (QW2) quantum wires (QWs), all of which show dynamics during the experiment. We also find that the edge of the encapsulated material appears amorphous and changes constantly under the electron beam, which complicates the identification of the elemental structure of the edge.
	
	\section*{Results}
	
	\noindent{\bf Atomic scale dynamics in free-standing MoTe$_2$}. As with other TMDCs~\cite{Zan2013, Nguyen2017, Iberi2016}, MoTe$_2$ monolayers damage quickly under electron irradiation (see Supplementary Figure 1). Nevertheless, it is possible to obtain atomic-resolution images of the structure when the exposure of the material is minimized. In Figure~1a we show an image sequence displaying initially a nearly perfect 1H structure, which is then partially converted into a 1T$'$ phase in panel a(v) (after an additional electron dose of $\sim 8.1\times 10^8~e^-/\mathrm{nm}^2$). The energy difference between 1H and 1T$'$ is just 0.03~eV/atom (see Supplementary Table 1) at zero pressure, and 1T$'$ becomes favored under strains between 0.3\% and 3\% at room temperature~\cite{Duerloo2014}. Hence, the local strain induced by the initially created Te vacancies (at a dose of $\sim 6.8\times 10^8~e^-/\mathrm{nm}^2$) most likely serves as the driving force for the observed transition from 1H to 1T$'$. After the vacancies have been created, some Mo atoms close to the defect shift to create diamond-shaped 4-membered rings. This leads eventually to the creation of a mixed 1H-1T$'$ area (after a dose of $\sim 8.3\times 10^8~e^-/\mathrm{nm}^2$). Further exposure to electron irradiation leads first to the creation of vacancies in the 1T$'$ phase and finally to its disappearance. Afterwards, only the defected 1H phase remains. After a dose of $\sim 12.5\times 10^8~e^-/\mathrm{nm}^2$, an 8-membered ring appears in addition to the diamond-shaped ones. At the end of the image sequence, a structure consistent with a three-fold rotationally symmetric (\textit{C}$_{3}$) defect made of 4-8-4-membered rings has been created.
	
	In another image sequence (Figure~\ref{gb}), we show a mirror-symmetric grain boundary between two 1H grains at $\sim$60$^\circ$ misorientation, similar to what has been reported in MoS$_2$, MoSe$_2$, and MoTe$_2$/MoS$_2$ heterostructures~\cite{Zhou2013,Hong2017,Diaz2016,Ma2017}. At least in our case, the area around the boundary was initially a single-crystalline area of 1H, and the misoriented grains only appeared during the experiment (see also Supplementary Figures 3, 4 and 5). During the image sequence (Figure~\ref{gb}), molybdenum atoms at the grain boundary (typically) undergo correlated migration by half a lattice vector (see for example the three atoms marked yellow in Figure~\ref{gb}a-i where the arrow indicates the direction of movement; the new positions are marked in panel (a-ii) with blue circles), allowing the migration of the grain boundary within the crystal. According to our density functional theory (DFT) simulations (Methods), the migration of the first Mo atom at the boundary has an energy barrier of 2.3 eV, dropping to 2.0 eV for the second one. The migration of the mirror-symmetric grain boundaries in MoTe$_{2}$ starts with the displacement of Mo atoms by half a lattice vector, which is similar to the grain boundary migration in MoSe$_{2}$~\cite{Lin2_2015}. In both materials, 
    chalcogen vacancies are formed in the lattice by electron irradiation before Mo atoms shift to stabilize the system. However, for the grain boundaries in WSe$_{2}$~\cite{Lin2015} and WS$_{2}$\cite{Azizi2014} the migration starts with the movements of chalcogen atoms around the dislocation core in contrast to MoTe$_{2}$ and MoSe$_{2}$.
	
	\begin{figure}[H]
		\centering
		\includegraphics[width=1.0\textwidth]{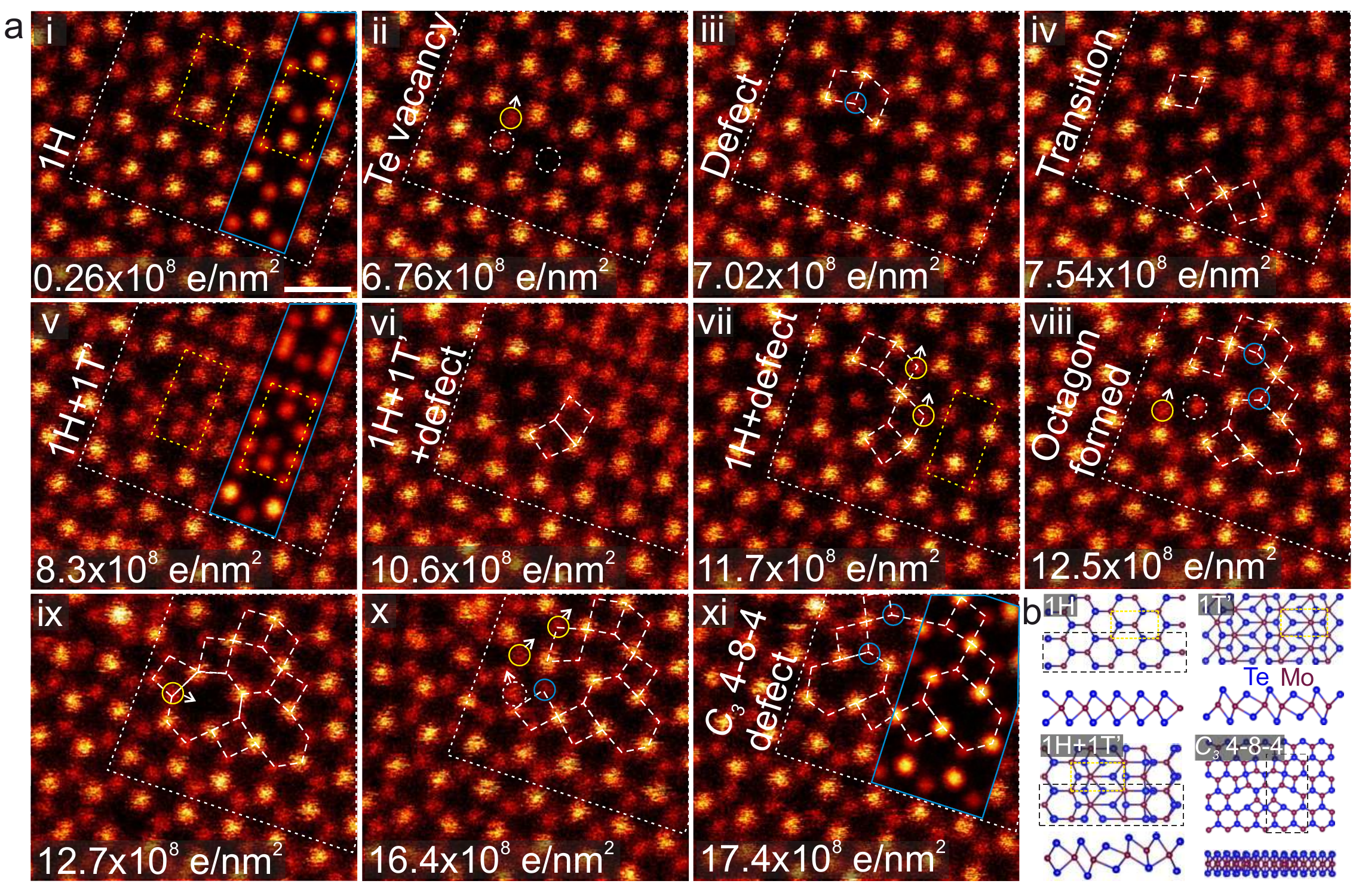}
		\caption{{\bf Phase transition and a three-fold rotationally symmetric defect in a MoTe$_2$ monolayer.} (a) STEM-HAADF image sequence showing two phase transitions (1H$\rightarrow$1T$'$ and 1T$'$$\rightarrow$1H) under electron irradiation. The overlaid areas outlined by blue lines in panels (a-i), (a-v) and (a-xi) are simulated images using the black dashed regions of the models in panel b. Images without overlays are shown in Supplementary Figure 2. The unit cells of the 1H and 1T$'$ phases are marked by yellow dashed frames. Te vacancies are marked by white dashed circles, while the Mo atom, before and after it shifts, is marked with yellow and blue solid circles, respectively. The areas surrounded by the white dashed lines show the same region in all frames. The experimental images display raw data in false color, and the scale bar is 0.5~nm.}
		\label{phasetrans}
	\end{figure}
	
	\begin{figure}[H]
		\centering
		\includegraphics[width=1.0\textwidth]{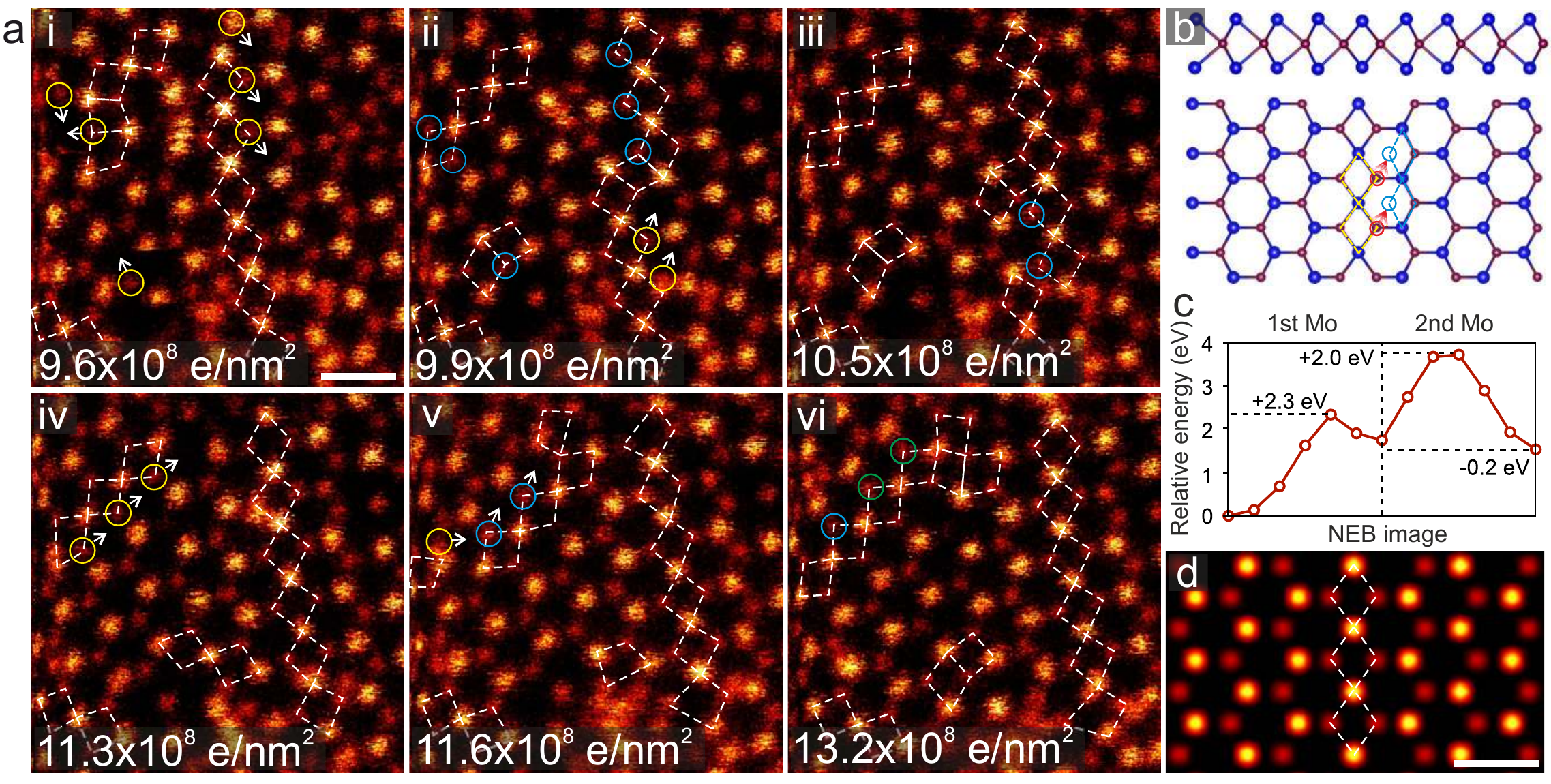}
		\caption{{\bf Migration of grain boundaries in a MoTe$_2$ monolayer.} (a) STEM-HAADF image sequence showing the migration of grain boundaries. The arrows on the images show the direction of movement of the Mo atoms marked with circles. (b) Simulated model of the grain boundary. The migration of the boundary atoms is schematically indicated with arrows, similar to the experimental images. (c) Migration barriers for two of the Mo atoms at a four-unit model of the grain boundary calculated via DFT using the nudged elastic band method. (d) Image simulation of the grain boundary model. The experimental images display raw data in false color, and the scale bars are 0.5~nm.} 
		\label{gb}
	\end{figure}
	
\begin{figure}[H]
		\centering
		\includegraphics[width=1.0\textwidth]{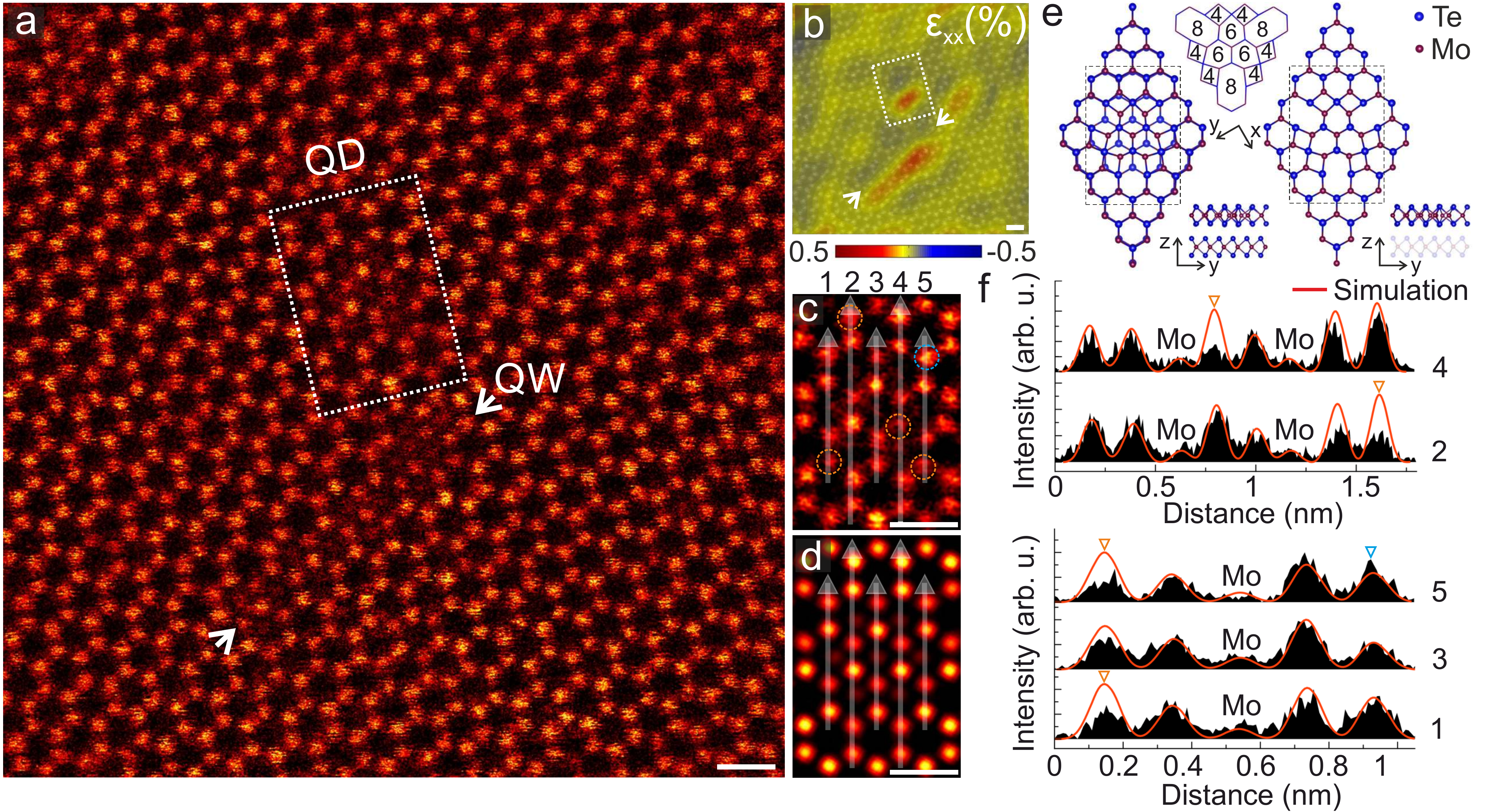}
		\caption{{\bf Structure of defects in encapsulated bilayer MoTe$_{2}$.} (a) STEM-HAADF image of an encapsulated MoTe$_{2}$ bilayer with a QD defect (white dashed rectangle) and QW (marked by arrows). Image background has been subtracted using a Gaussian blur with a radius of 20~px. (b) $\varepsilon_{xx}$ strain map of panel (a) overlaid on the atomic structure. (c) A close-up of the QD (treated by a Gaussian blur with a radius of 1~px). Orange and blue dashed circles show atomic vacancies and additional atoms, respectively. (d) Image simulation for the atomic model shown in panel (e). (e) QD model with and without the second MoTe$_{2}$ layer. Top views of the QD atomic model with a three-fold rotationally symmetric defect consisting of 4-8-4 membered rings in the upper MoTe$_{2}$ layer. (f) Intensity profiles along the white semi-transparent arrows shown in panels (c) and (d). Triangles show the locations of atomic vacancies and additional atoms as indicated in panel (c). The scale bars are 0.5~nm.}
		\label{QD}
	\end{figure}

	
	\noindent{\bf Quantum dots.} One of the most commonly observed defects in our samples is a QD structure shown for an encapsulated bilayer in Figure~\ref{QD} (graphene is not visible due to the low contrast of C atoms as compared to Mo and Te; see also Supplementary Figure 6). We observed a concentration of 0.58/100~nm$^{2}$ for the QDs (total area of 9500~nm$^{2}$ was imaged). This defect appears as a ca. 1~nm round feature within the 2H phase. Because of the different atomic structure with respect to the surrounding crystal, the QD can also be visualized through geometric phase analysis (along with the resulting strain maps, see Figure~\ref{QD}b). Within the QD, the atomic structure has the appearance of the 1T phase. To ascertain the actual atomic structure, we created a large number of candidate structures for the defects and simulated STEM images based on DFT-relaxed atomic configurations. The best match to the experimental images was obtained for the structure shown in Figure~\ref{QD}. In this structure, one of the layers of the bilayer MoTe$_{2}$ involves a three-fold rotationally symmetric defect consisting of 4-8-4 membered rings whereas the other layer is pristine. This defect is similar to that found in the non-encapsulated material after extensive electron-beam irradiation (Figure~\ref{phasetrans}a-xi). An atomic model of the defected layer is shown in panel (e). Intensity profiles recorded on the white lines over the simulated (panel (d)) and experimental images (panel (c)) show that the atomic model and the experimental image are in a good agreement. The intensity profiles also reveal that there are Te vacancies and one additional atom in this experimental QD structure as compared to the model (such irregularities, appearing at different atoms for each defect, are common among all the QD structures we observed). Unlike defects created by electron irradiation, these defects in one layer of the encapsulated MoTe$_{2}$ bilayer are likely to have been formed during sample growth or preparation since they were observed from the beginning of the imaging and formation of additional QDs was only very rarely observed during the experiments. 
	
\begin{figure}[H]
		\centering
		\includegraphics[width=1.0\textwidth]{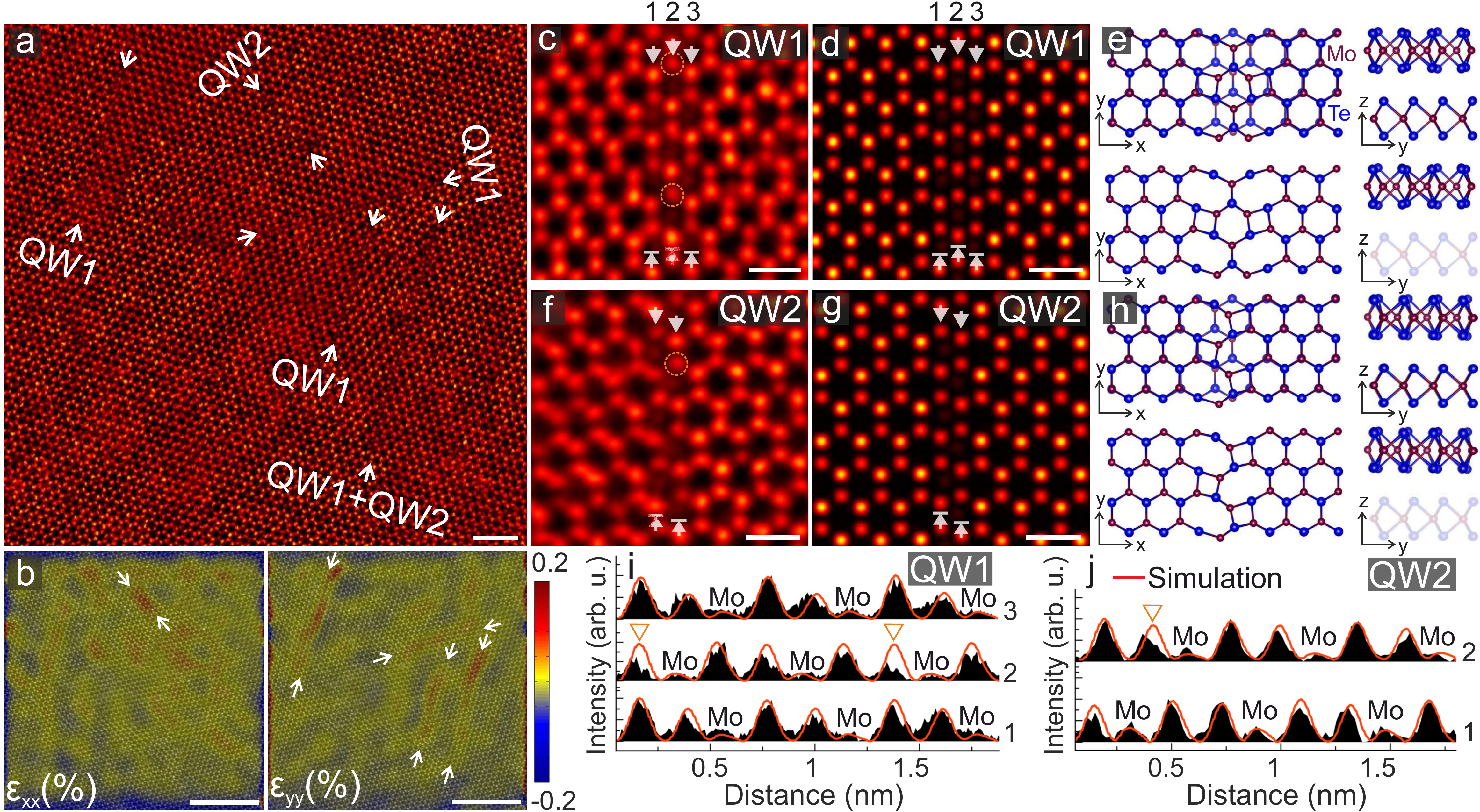}  
		\caption{{\bf Atomic structure of quantum wires.} (a) STEM-HAADF image of several QWs in encapsulated bilayer MoTe$_{2}$. Image background has been subtracted using a Gaussian blur with a radius of 12~px. The scale bar is 2~nm. (b) $\varepsilon_{xx}$ and $\varepsilon_{yy}$ strain maps of panel (a). The scale bar is 4~nm. (c,f) Close-up images of the two different QW structures (treated by a Gaussian blur with a radius of 3~px). Orange dashed circles show atomic vacancies. (d,g) Simulated images corresponding to the experimental structures. The scale bars are 0.5~nm. (e,h) Atomic models of the simulated images (top: both layers, bottom: only the defective layer). (i,j) Intensity profiles along the white arrows on the experimental and simulated QW images. Triangles show the locations of atomic vacancies as indicated in panels (c) and (f).}
		\label{QW}
	\end{figure}

	\noindent{\bf Quantum wires.} Besides the QD structures, our encapsulated samples also contain two different kinds of QWs (one was visible in Figure~\ref{QD}a). Several examples of both are shown in Figure~\ref{QW}. The concentration of QWs (0.48/100~nm$^{2}$ in a total observed area of 9500~nm$^{2}$) is slightly lower than that of the QDs, and they can also be easily detected from microscopy images through strain maps (Figure~\ref{QW}b), either in $\varepsilon_{xx}$ or $\varepsilon_{yy}$ depending on the direction of the QW with respect to the host lattice. A close-up image of the first type (QW1) is shown in Figure~\ref{QW}c along with a simulated image (Figure~\ref{QW}d) and the corresponding atomic structure (Figure~\ref{QW}e); QW2 is similarly displayed in Figure~\ref{QW}f-h. The atomic structures of the defects are shown in Figure~\ref{QW}e,h. QW1 includes reflection symmetric ($\sigma{_v}\|$) units consisting of 8-4 membered rings (Figure~\ref{QW}e). In contrast, QW2 is made of two-fold rotational symmetric units (\textit{C}$_{2}$) also consisting of 8-4 membered rings (Figure~\ref{QW}h). Line profiles over the experimental and simulated STEM-HAADF images are plotted in Figure~\ref{QW}i,j. Similarly to QDs, QW structures also include Te vacancies. QD and QW defects are less stable compared to pristine 1H-MoTe$_{2}$ and their formation energies are found to be 2.80, 4.81 and 3.73 eV respectively for QD, QW1 and QW2 (see also Supplementary Table 2). Among these defect structures, QD has one more Mo atom than a pristine structure of the same size, whereas the QWs are stoichiometric. The formation energy of the QD has thus been calculated using the chemical potential for bulk molybdenum. Line defects in the armchair direction were previously reported for few-layer MoTe$_{2}$~\cite{Zhao2017}. In that work, it was concluded that line defects are mostly observed in only one of the MoTe$_{2}$ layers within the few-layer structures, similar to the defects reported here. Further, it was suggested that line defects are formed via the rearrangement of Te vacancies. By contrast, the line defects discussed in our study are formed via the rearrangement of Mo atoms.
	
\begin{figure}[H] 
		\centering
		\includegraphics[width=1.0\textwidth]{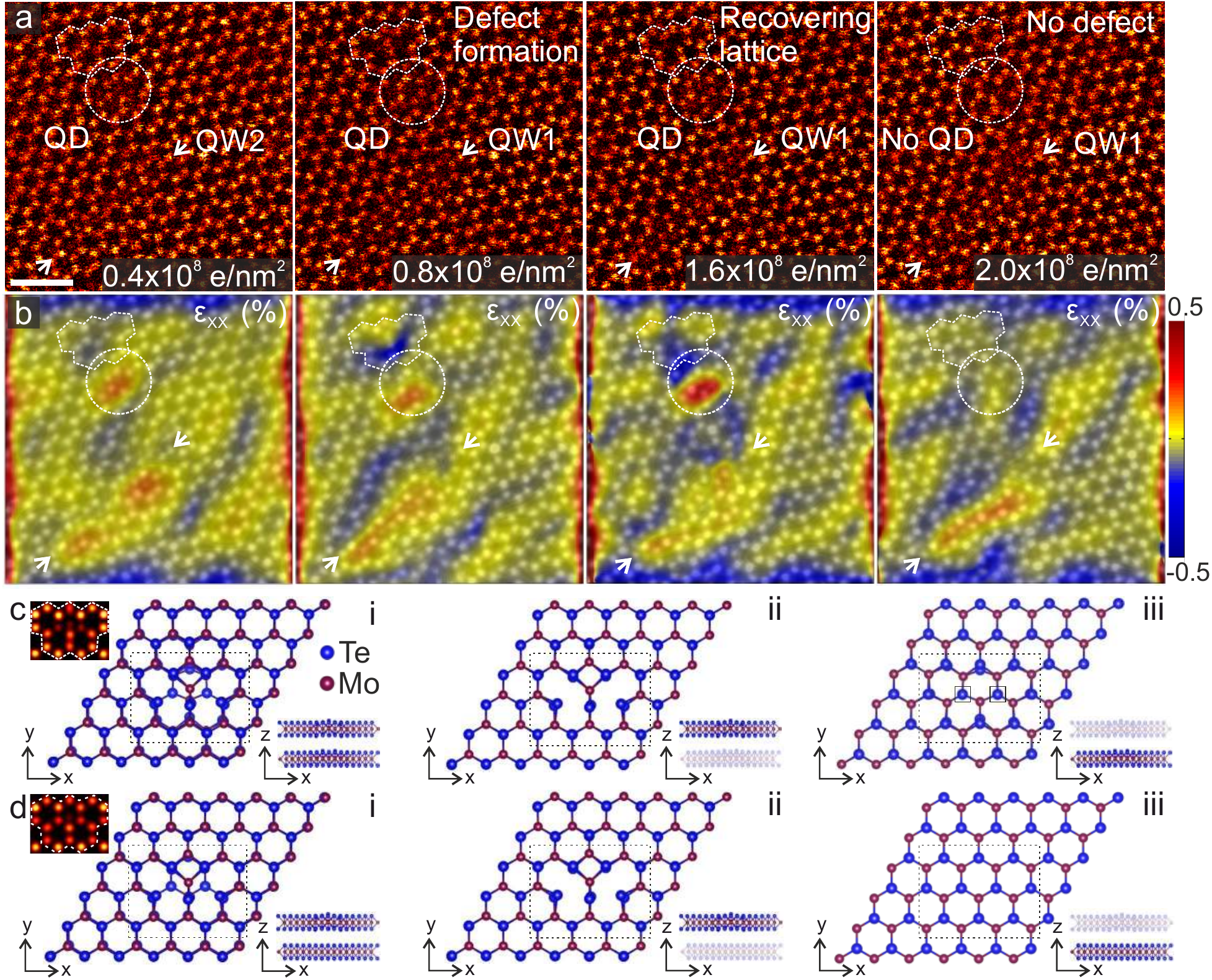}
		\caption{{\bf Dynamics of QDs and QWs under electron irradiation.} (a) STEM-HAADF image sequence showing the change of the atomic structure of a QD and QW in encapsulated bilayer MoTe$_{2}$ (background subtracted using a Gaussian blur of 20~px). The scale bar is 0.5~nm. (b) $\varepsilon_{xx}$ strain maps corresponding to panel (a). (c,d) Atomic models of the butterfly-like defect in the second frame of panel (a). The areas marked by dashed black frames correspond to those shown in simulated images in the insets. The black squares show the location of Te vacancies at the second MoTe$_{2}$ layer shown in panel (c-iii).}
		\label{dynamics}
	\end{figure}

	\noindent{\bf Evolution of QD and QW structures during observation.} Figure~\ref{dynamics} shows an image sequence of an area containing both a QD and a QW (initially having the QW2 structure). Strain maps corresponding to the image sequence in Figure~\ref{dynamics}a are shown in Figure~\ref{dynamics}b. Although the QW is barely visible in these strain maps due to its orientation, the QD is easily visible. After an additional electron dose of $\sim 0.4\times 10^8 ~e^-/\mathrm{nm}^2$, a vacancy defect appears next to the QD. At the same time, presumably due to the negative strain (increasing blue area), the QW structure is transformed from a two-fold rotation-symmetric type (QW2) to a reflection-symmetric (QW1) type. Although the defect structure in the second frame is a bit unclear, it appears to involve one slightly displaced Mo atom, two missing Te and one missing Mo, as shown in panel (c-i). To further understand this structure, we created two possible atomic models, as shown in panels (c,d). While the first layer in panel (c-ii) involves one shifted and one missing Mo atom, the second layer in panel (c-iii) has two missing Te atoms. The other model (panel (d-i)) has no missing Te atom at the second layer. From the simulated images of both models (see insets of panels (c-i) and (d-i)), we conclude that the structure with two missing Te atoms at the second MoTe$_{2}$ layer is in better agreement with the experimental image. In the third frame of panel (a), the displaced Mo atom has moved back to its lattice position. Finally, in the last frame, both Mo and Te vacancies are filled and at the same time the QD disappears. Another example of atomic dynamics at a QD is shown in Supplementary Figure 7.
	
	\noindent{\bf Electronic structure.} Since both of the observed defect types are created within the semiconducting 1H phase, it is a worth asking whether their electronic properties differ significantly from those of the host structure. The electronic band structures calculated for the 1H phase both in the primitive hexagonal 3-atom and orthorhombic 6-atom unit cells are shown in Figure 6a and d. The DFT band gap is ca. 1.14~eV. The unfolded electronic band structures for the QD (hexagonal supercell) and QWs (orthorhombic supercells) are shown in panels b, e and f. While the extra bands due to the QD defect resemble localized molecular states, in both the QW1 and QW2 defects there is clear evidence for metallic conduction in the direction of the defect line. This suggests the possibility of applications as conductive wires or nanoscale antennas.
	
	\begin{figure}[H]
		\centering
		\includegraphics[width=1.0\textwidth]{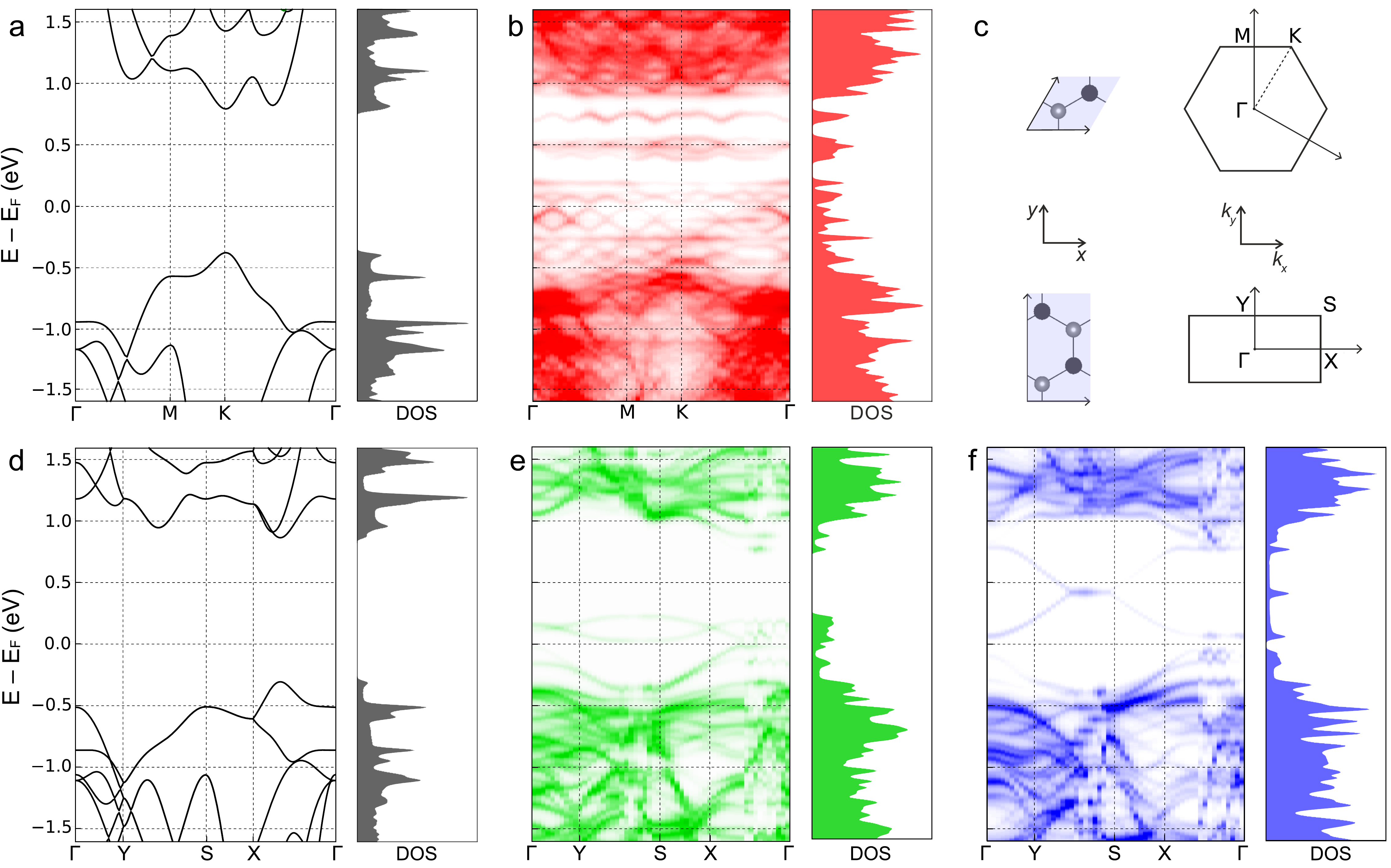}
		\caption{{\bf Electronic structure calculations.} Unfolded electronic band structures calculated with DFT. The band structure (hexagonal cell) of the (a) 1H phase of monolayer MoTe$_{2}$ and (b) QD defect. (c) The lattice and the first Brillouin zone of the 3-atom and 6-atom unit cells. The electronic band structure (orthorhombic cell) of the (d) 1H phase of monolayer MoTe$_{2}$, (e) QW1 defect and (f) QW2 defect. The intensity of some of the bands (in panels b, e and f) is weaker because these states are due to the localized defect in the unfolded band structure, and the segmentation of the lines reflect the finite k-point sampling.}
		\label{electronic}
	\end{figure}
	
	
	\noindent{\bf Edge structure.} Finally, despite the overall stability of the encapsulated MoTe$_2$ under the electron beam, its edges appear unstable. Figure~\ref{edge}a shows the atomic structure of an edge in bilayer MoTe$_{2}$. Ideal symmetric edges would be terminated either by Te or Mo atoms. However, elemental identification at the edge is complicated by the liquid-like behavior of the structure during continuous imaging (see Supplementary Videos and Supplementary Figure 8). This is in contrast to the non-encapsulated structure, for which the atomic structure remains ordered (see Supplementary Figure 9). To elucidate on the disordered structure, we carried out molecular dynamics simulations of an encapsulated MoTe$_2$ structure at room temperature. These simulations reveal a clear tendency for the Te atoms to detach from the MoTe$_2$ crystal and spread between the graphene sheets. Interestingly, no electron-beam effect is required to produce this behavior. Due to the similarity between the experimental image (Figure~\ref{edge}b) and the simulated one (Figure~\ref{edge}c) we believe that Te atoms are also responsible for the dynamics observed in the microscope. Following Ref.~\cite{Vasu2016}, we estimate the pressure due to graphene encapsulation to be in the order of a few GPa close to the edge of the MoTe$_2$ structure. In contrast to our present results with MoTe$_2$, graphene-encapsulated MoS$_2$ has been reported to have an easily resolvable edge structure~\cite{Siller2013,Zan2013}.  
	
	\begin{figure}[H]
		\centering
		\includegraphics[width=1.0\textwidth]{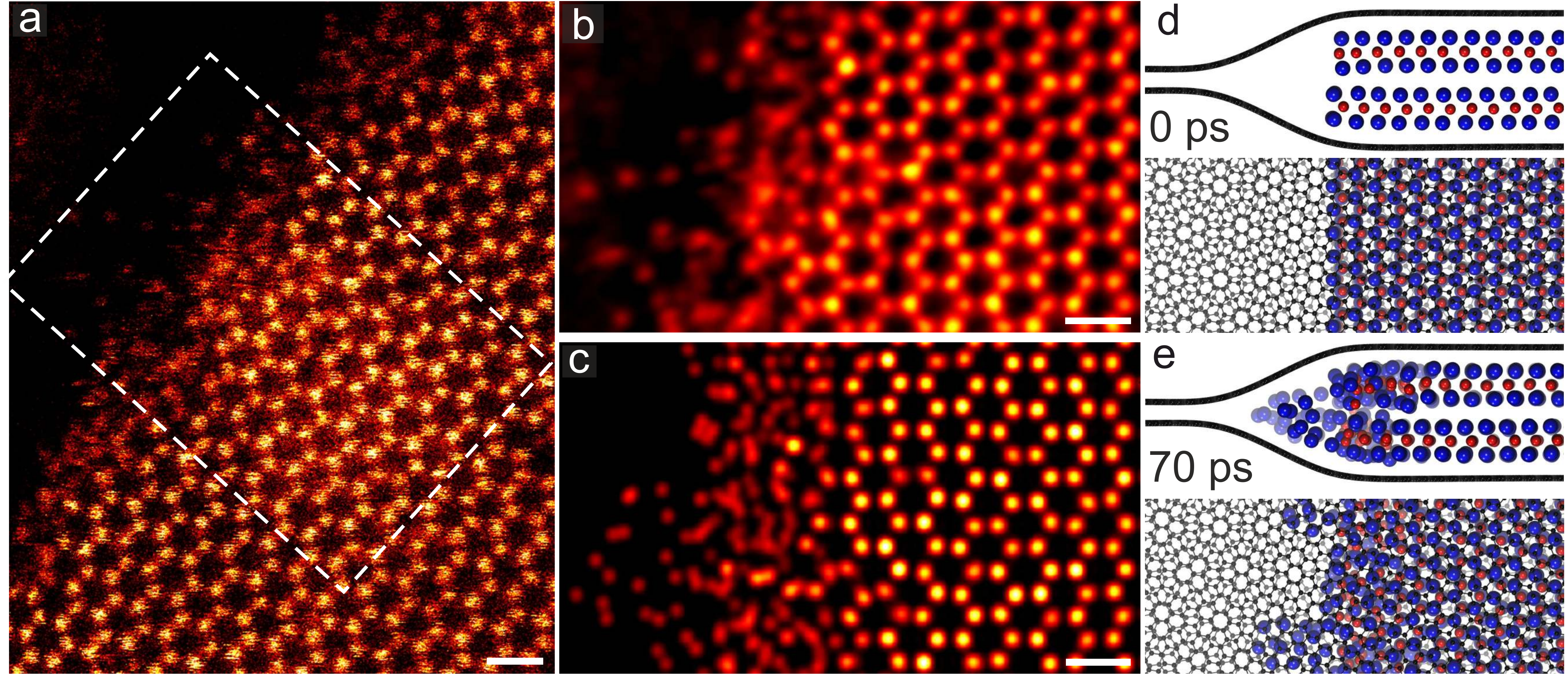}
		\caption{{\bf Edge structure of encapsulated bilayer MoTe$_{2}$.} (a) STEM-HAADF image showing the edge of graphene-encapsulated bilayer MoTe$_{2}$. The image displays raw data in false color. (b) A close-up of the region indicated by the white dashed line in panel (a). The image has been treated by a Gaussian blur with a radius of 4~px. (c) Simulated snapshot of graphene-encapsulated bilayer MoTe$_{2}$ after 70~ps at 300~K. The scale bars are 0.5~nm. (d,e) Side and top views of the model structure at the beginning of the simulation and after 70~ps at 300~K.}
		\label{edge}
	\end{figure}
	
	
	\section*{Conclusions}
	
	Unprotected MoTe$_2$ is highly susceptible to damage during TEM imaging, which leads to the creation of vacancies, especially at the edges of the suspended material, but also to interesting atomic-scale dynamics. For example, the semiconducting 1H phase can be locally turned into the semi-metallic 1T$'$ phase, which can in turn be transformed back to defective 2H due to continuing loss of atoms. Electron irradiation also leads to the migration of grain boundaries between 1H grains with a $60^\circ$ misorientation. Further, a three-fold rotationally symmetric defect appears at high electron doses. All of these dynamics are significantly suppressed when MoTe$_2$ is encapsulated between two graphene layers, which allows the study of its intrinsic defects. We observe a relatively high concentration (0.58/100~nm$^2$) of 1-nm-sized quantum dot structures, associated with the three-fold rotationally symmetric defect, as well as two different quantum wire structures (with a joint concentration of about 0.48/100~nm$^2$): a reflection-symmetric defect (QW1) and two-fold rotationally symmetric defect (QW2). Both the possibility of introducing phase transitions between electrically very different phases and the existence of quantum dot and quantum wire defects with midgap states in the otherwise semiconducting material indicate a possibility for tuning the material properties through defect engineering. The amorphous nature of the edge in encapsulated MoTe$_2$ can lead to transport of Te atoms through the edge, and thus a metallic Te ion channel surrounding the semiconducting MoTe$_2$ can be obtained.
	
	
	\section*{Methods} 
	\setlength{\parskip}{1em}	
	\noindent{\bf Sample preparation.} The MoTe$_{2}$ flakes were exfoliated on a SiO$_{2}$/Si substrate and then transferred onto TEM grids with a carbon-based support membrane (QUANTIFOIL\textsuperscript{\textregistered}). Monolayer graphene sheets were synthesized by chemical vapor deposition (CVD) in a mixture of 50~sccm CH$_{4}$ and 2000~sccm Ar/H$_{2}$ gases at 960$^\circ$C. First, a graphene sheet was transferred from Cu foil onto a TEM grid without polymer, and the MoTe$_{2}$ flakes then transferred on top. The obtained MoTe$_{2}$/graphene stack was then transferred onto another graphene sheet on Cu foil. The Cu foils were etched with FeCl$_{3}$.
	
	\noindent{\bf Transmission electron microscopy.} STEM images of graphene-encapsulated MoTe$_{2}$ were recorded using a Nion UltraSTEM 100 electron microscope operated at a 60~kV accelerating voltage in near ultra-high vacuum ($2\times 10^{-7}$~Pa) using the high angle annular dark field (HAADF) detector with a collection angle of $80-200$~mrad. For TEM diffraction experiments (Supplementary Material) we used a Delong instruments LVEM5 table-top transmission electron microscope operated at 5~kV.
	
	\noindent{\bf DFT and STEM image simulations.} Density functional theory (DFT) simulations were carried out using the grid-based projector-augmented wave (GPAW) software package \cite{Mortensen2005} to study the properties of the unit cell and supercells of monolayer and bilayer MoTe$_{2}$. For unit cells, we used a plane-wave basis (cutoff energy 700~eV, $16 \times 16 \times 1$ \textbf{k}-point mesh) to relax the atomic positions using the C09 van der Waals functional~\cite{Cooper2010} (the DF2 functional~\cite{Lee2010} was additionally used for relative energies of different MoTe$_{2}$ phases shown in Supplementary Table 1). The atomic structure of supercells of MoTe$_{2}$ including QDs and QWs were relaxed with periodic boundary conditions in the finite-difference mode with the grid spacing of 0.2 \AA{} and a $3 \times 3 \times 1$ \textbf{k}-point mesh so that maximum forces were $<$0.05 eV \AA{}$^{-1}$. A double-zeta linear combination of atomic orbitals basis was used to speed up the calculations for the larger simulated structures. STEM-HAADF and TEM diffraction simulations were performed using the QSTEM software with parameters corresponding to the experiments~\cite{Koch2002}. The NEB simulation was performed with no frozen atoms and with five intermediate images~\cite{Henkelman2000}. All optimized atomic structures are provided in the Supplementary Information.
	
	\noindent{\bf Molecular dynamics.} To study the atomic structure of MoTe$_2$ edges encapsulated by graphene, we created a supercell with a size of 28$\times$58~{\AA}$^2$ of bilayer 2H MoTe${_2}$. The two graphene sheets are misoriented by $\sim$18$^\circ$ with respect to each other, and the resulting moir\'{e} pattern periodicity is $\sim$14~{\AA}. The reactive bond-order-dependent force field that supports bond breaking and bond formation was used to describe the interaction between molybdenum and tellurium atoms and also the interaction between the carbon atoms~\cite{Aktulga2012245,Onofrio2016}. Long distance van der Waals interactions between the carbon atoms and between C-Mo and C-Te were treated with Morse potentials. All calculations were performed with Large-scale Atomic/Molecular Massively Parallel Simulator (LAMMPS) code~\cite{PLIMPTON19951,Plimpton2012}. The total potential energy was minimized by relaxing both layers without applying any constraints until the forces were below $10^{-3}$~eV/{\AA} and the strain on the whole structure negligible (pressure below 1~bar). Later, the temperature was increased to 300~K for 1~ns for the molecular dynamics simulations (graphene sheets were kept at 0~K).
	
	\section*{Supporting Information} 
	
	TEM and electron diffraction images, STEM-HAADF images of monolayer, bilayer and encapsulated MoTe${_2}$, the atomic model of graphene-encapsulated MoTe${_2}$ monolayer, DFT total energies of 2H, 1T, and 1T$'$ phases of MoTe$_{2}$ monolayers and bilayers, formation energies of monolayer MoTe$_{2}$ defects, atomic configuration files of the structures shown in the manuscript and videos showing the edge of graphene-encapsulated MoTe${_2}$ bilayer. 
	
	\section*{Acknowledgements} 
	K.E., M.R.A.M. and J.C.M. acknowledge support from the Austrian Science Fund (FWF) through project P25721-N20, T.S. through project P~28322-N36, and M.R.A.M. and J.K. through project I3181-N36. T.J.P. was supported by the European Union's Horizon 2020 research and innovation programme under the Marie Sk\l odowska-Curie grant agreement No. 655760--DIGIPHASE. J.C.M. acknowledges funding by the European Research Council Grant No. 336453-PICOMAT and J.K. from the Wiener \mbox{Wissenschafts-,} Forschungs- und Technologiefonds (WWTF) via project MA14-009. We also acknowledge generous computational resources from the Vienna Scientific Cluster.
	
	\bibliography{bibliography}

\newpage

\begin{center}
    \huge{\bf Supplementary Information}
\end{center}

	\begin{figure}[H]
		\renewcommand\figurename{Supplementary Figure}
		\setcounter{figure}{0}
		\centering
		\includegraphics[width=0.75\textwidth]{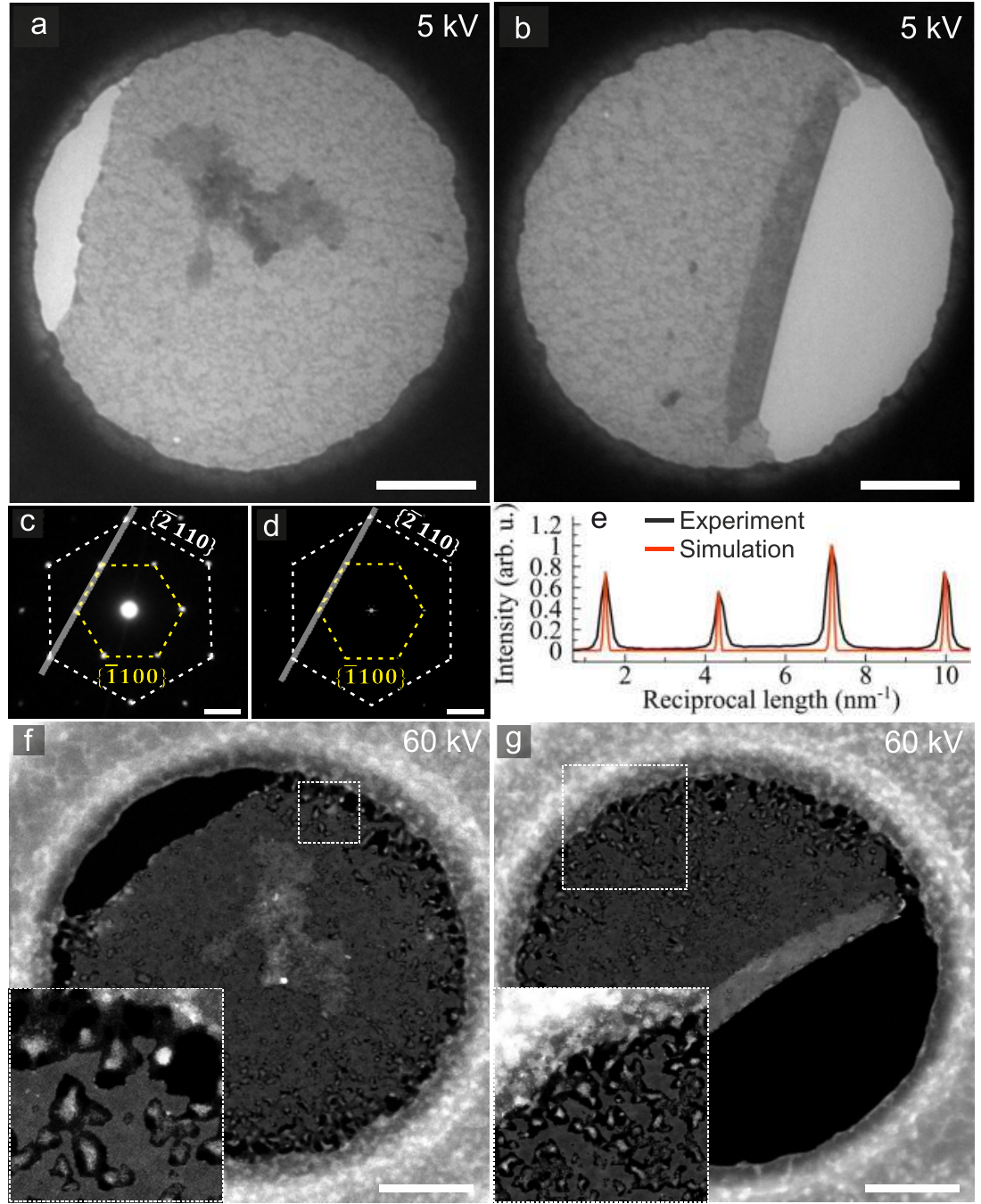}
		\caption{Electron beam damage in the MoTe$_{2}$ monolayer. (a,b) TEM images recorded at 5 kV. (c,d) Experimental and simulated diffraction patterns of monolayer MoTe$_{2}$. (e) Intensity profiles along the white semi-transparent lines over the \{$\overline{1}100$\} and \{$\overline{2}110$\} families of experimental and simulated diffraction patterns. (f,g) STEM-HAADF images of the same sample recorded at 60 kV. Insets are close-ups corresponding to the regions shown by white dashed frames. The scale bars on the TEM and STEM images are 200 nm and on the diffraction patterns 2~nm$^{-1}$.}
		\label{SIFigure 1}
	\end{figure}
	\begin{figure}[H]
		\renewcommand\figurename{Supplementary Figure}
		\centering
		\includegraphics[width=1.0\textwidth]{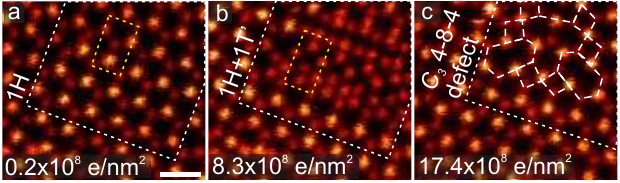}
		\caption{STEM-HAADF images of MoTe$_{2}$ monolayer without the overlays. STEM-HAADF images shown in panel (a), (b), and (c) are the same images shown in Figure~\ref{phasetrans}i, v, and xi, respectively. The unit cells of the 1H and 1T$'$ phases are marked by yellow dashed frames. All images display raw data in false color, and the scale bar is 0.5~nm.}
		\label{SIFigure 2}
	\end{figure}
	\begin{figure}[H]
		\renewcommand\figurename{Supplementary Figure}
		\centering
		\includegraphics[width=1.0\textwidth]{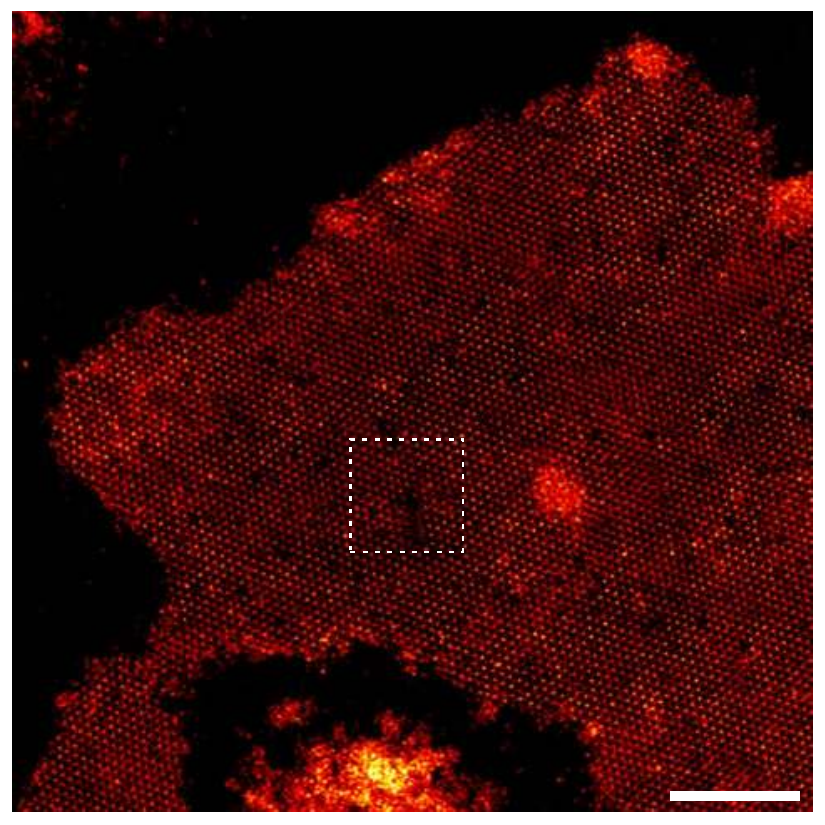}
		\caption{An overview STEM-HAADF image of monolayer MoTe$_{2}$. The dashed frame shows the region where the GB images shown in Figure 2 of the main manuscript have been recorded. The scale bar is 5 nm.}
		\label{SIFigure 3}
	\end{figure}
	\begin{figure}[H]
		\renewcommand\figurename{Supplementary Figure}
		\centering
		\includegraphics[width=1.0\textwidth]{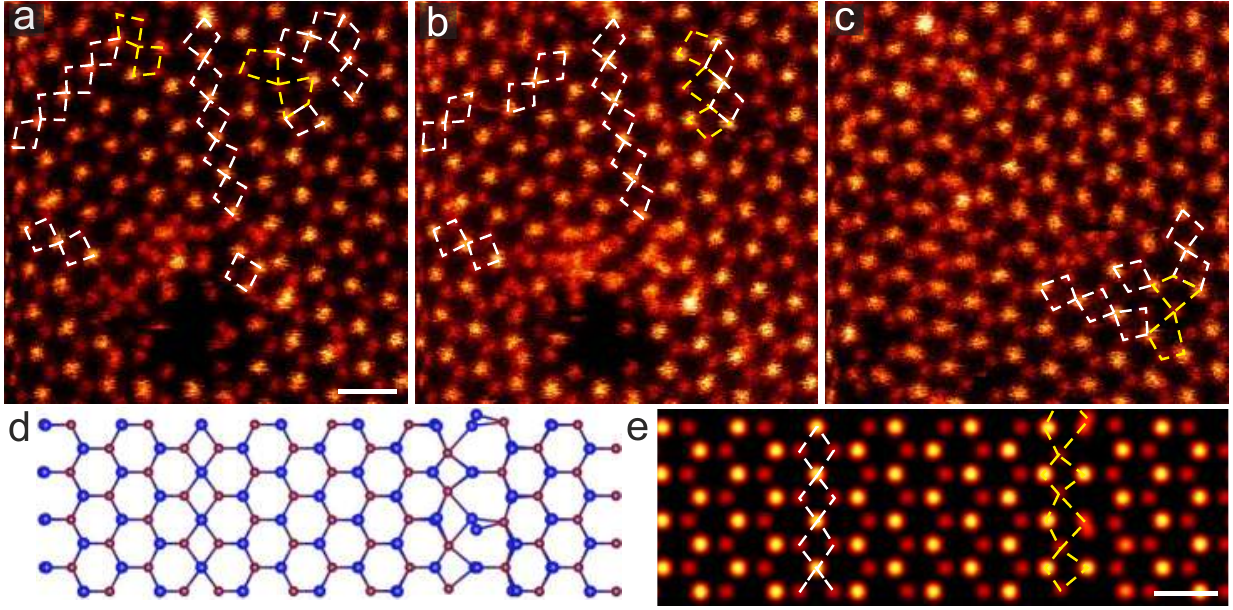}
		\caption{(a-c) STEM-HAADF images of the MoTe$_{2}$ monolayer showing two different polarities of mirror-symmetric grain boundaries indicated by the white and yellow dashed diamonds. (d) DFT-relaxed atomistic model of monolayer MoTe$_{2}$ with grain boundaries. (e) Corresponding simulated STEM-HAADF image of the model. The scale bars are 0.5 nm.} 
		\label{SIFigure 4}
	\end{figure}
	\begin{figure}[H]
		\renewcommand\figurename{Supplementary Figure}
		\centering
		\includegraphics[width=1.0\textwidth]{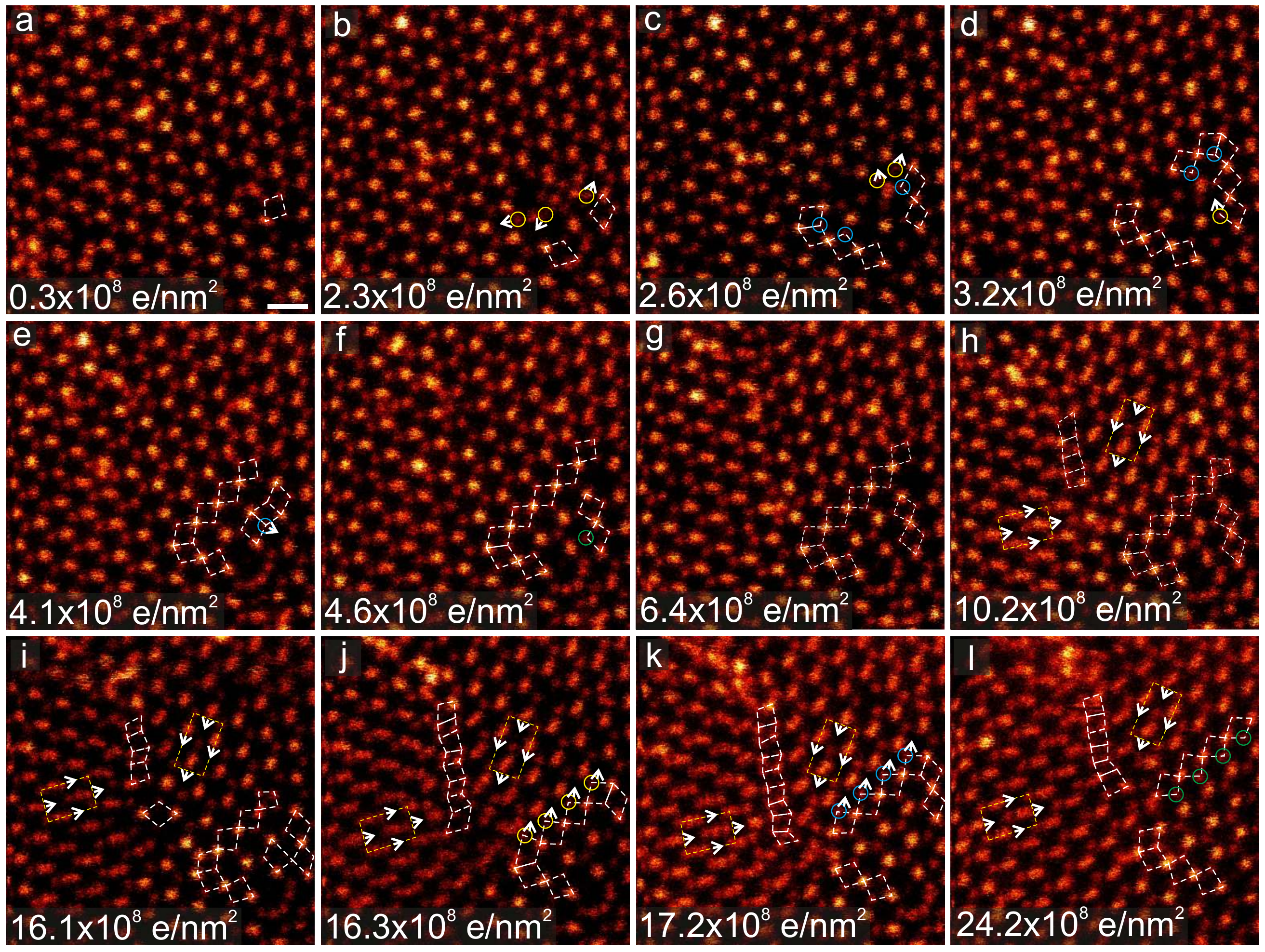}
		\caption{STEM-HAADF images indicating the formation and migration of mirror-symmetric grain boundaries during continuous electron exposure. The white arrows on the images show the direction of movement of the Mo and Te atoms. The initial positions of Mo atoms before they move are shown by yellow  circles while their second and third positions are respectively shown by blue and green circles. The unit cell of 1T$'$ MoTe$_{2}$ is indicated by the yellow dashed frames. The shift in Te atoms starts after electron dose of $\sim 10.2\times 10^8~ e^-/\mathrm{nm}^2$ (see panels (h-l)). All images display raw data in false color. The scale bar is 0.5 nm.} 
		\label{SIFigure 5}
	\end{figure}
	\begin{figure}[H]
		\renewcommand\figurename{Supplementary Figure}
		\centering
		\includegraphics[width=1.0\textwidth]{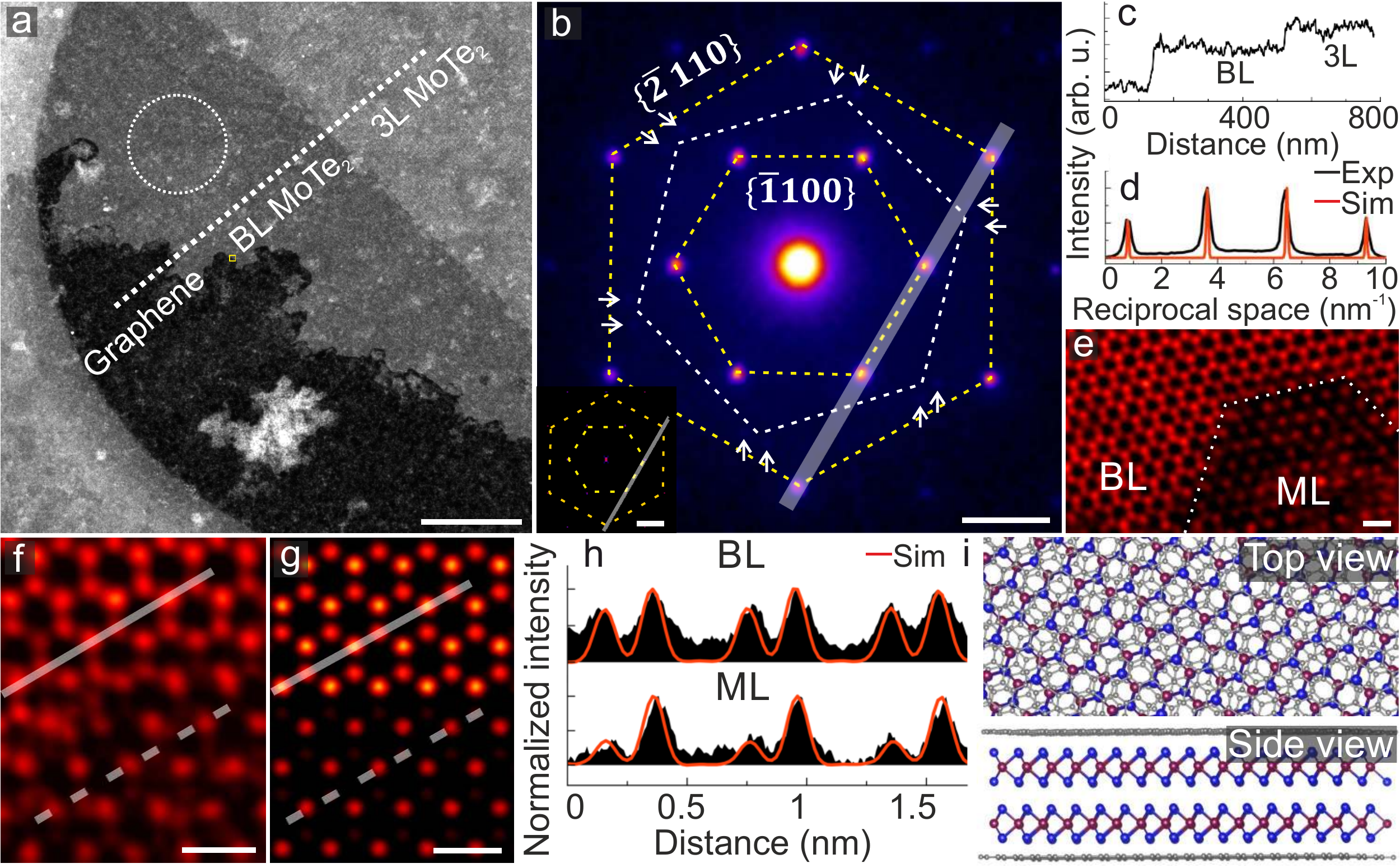} 
		\caption{ (a) STEM-HAADF overview image of graphene-encapsulated MoTe$_{2}$. The white dashed line indicates the position of the line profile in panel (c). The scale bar is 200 nm. (b) TEM diffraction pattern of the heterostructure from the 200~nm area that is shown with the white dashed circle in (a). The diffraction spots of graphene are on the white dashed line, with the white arrows indicating the two graphene layers. The diffraction spots of MoTe$_{2}$ are much brighter than those of graphene and are marked by the yellow dashed lines. Inset shows the simulated diffraction pattern of MoTe$_{2}$ bilayer. The scale bars are 2 nm$^{-1}$. (c) Intensity profile along the white dashed line on panel (a). (d) Intensity profiles of experimental and simulated diffraction patterns along the white semi-transparent lines over the \{$\overline{1}100$\} and \{$\overline{2}110$\} families for experimental and simulated TEM diffraction spots shown in panel (b) and its inset. (e) Atomic resolution STEM-HAADF image of the graphene-encapsulated monolayer and bilayer MoTe$_{2}$, recorded at the region indicated by the small yellow frame in panel (a). The scale bar is 0.5 nm. (f,g) Close-up STEM-HAADF image of the graphene-encapsulated MoTe$_{2}$ and the corresponding simulated HAADF image. The HAADF images have been treated by a Gaussian blur with a radius of 4~px. The scale bars are 0.5 nm. (h) The intensity profiles along the white solid (on bilayer) and dashed (on monolayer) semi-transparent lines in panels (f,g). (i) A schematic illustration of top and side views of the model used in the image simulations.} 
		\label{SIFigure 6}
	\end{figure}
	\begin{figure}[H]
		\renewcommand\figurename{Supplementary Figure}
		\centering
		\includegraphics[width=1.0\textwidth]{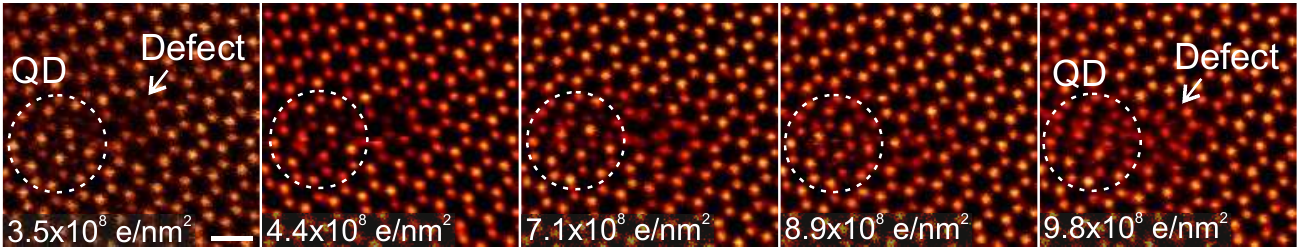}
		\caption{Atomic resolution STEM-HAADF images showing the structure of the QD embedded in bilayer MoTe$_{2}$ under different electron doses. The background of the image has been subtracted using a Gaussian blur with a radius of 32~px. The scale bar is 0.5 nm.}
		\label{SIFigure 7}
	\end{figure}
	\begin{figure}[H]
		\renewcommand\figurename{Supplementary Figure}
		\centering
		\includegraphics[width=1.0\textwidth]{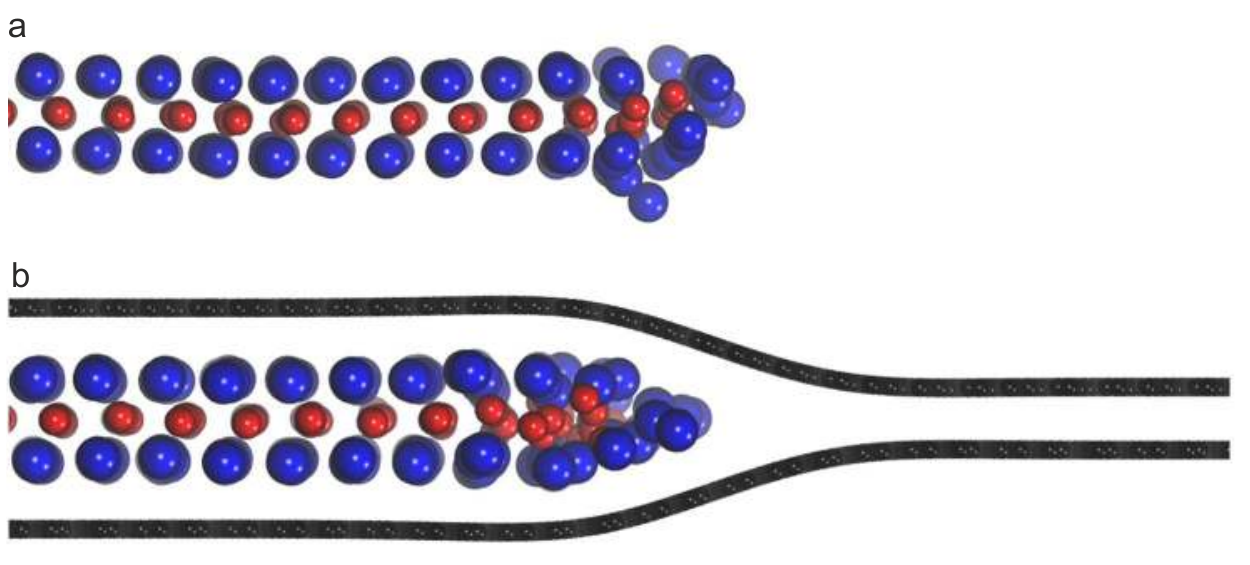}
		\caption{(a) MoTe$_{2}$ monolayer and (b) graphene-encapsulated MoTe$_{2}$ monolayer simulated by molecular dynamics at 300 K after 70~ps.}
		\label{SIFigure 8}
	\end{figure}
		\begin{figure}[H]
		\renewcommand\figurename{Supplementary Figure}
		\centering
		\includegraphics[width=1.0\textwidth]{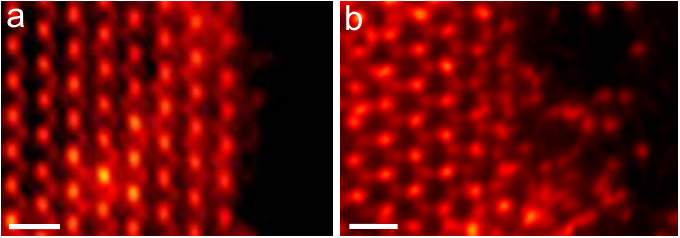}
		\caption{STEM-HAADF images of the edge of (a) the non-encapsulated and (b) graphene-encapsulated MoTe$_{2}$ monolayer. The scale bars on the images are 0.5 nm. The images have been filtered using a Gaussian blur with a radius of 3~px.}
		\label{SIFigure 9}
	\end{figure}
	\begin{table}[H]
		\renewcommand\figurename{Supplementary Figure}
		\centering
		\caption{Relative DFT total energies of 2H, 1T, and 1T$'$ phases of MoTe$_{2}$ monolayers (ML) and bilayers (BL) for two different van der Waals functionals (DF2-vdW and vdw-C09). Since the unit cell of 1T$'$-BL MoTe$_{2}$ has 6 atoms compared to 2H and 1T phases having 3 atoms, the DFT relative energies of ML and BL 1T$'$-BL MoTe$_{2}$ are divided by two to directly compare the relative energies of different phases. All values are given in units of eV/cell.}
		\label{tab:2}  
		\begin{tabular} {|c|c|c|}
			\hline
			Structure & DF2-vdW & vdW-C09\\
			\hline\hline
			1H-ML & 0 & 0 \\
			\hline
			2H-BL & -0.083 & -0.171 \\
			\hline
			1T-ML & 0.495 & 0.500 \\
			\hline
			1T-BL & 0.401 & 0.421 \\
			\hline
			1T$'$-ML & 0.013 & 0.034 \\
			\hline
			1T$'$-BL & -0.037 & -0.031 \\
			\hline
		\end{tabular} 
	\end{table}
	\begin{table}[H]
		\renewcommand\figurename{Supplementary Figure}
		\centering
		\caption{Formation energies of monolayer MoTe$_{2}$ defects (E$_{f}$) are calculated using the PBE functional with respect to supercells of pristine MoTe$_{2}$ of the same size. The chemical potential calculated for bulk molybdenum ($\mu$$_{Mo}$) is used for calculating the formation energy of the QD, as it has one more Mo atom than the pristine structure. All values are given in units of eV/cell.}
		\label{tab:1}  
		\begin{tabular} {|c|c|c|c|c|}
			\hline
			Structure &E$_{pristine}$ & $\mu$$_{Mo}$ & E$_{defect}$ & E$_{f}$\\
			\hline\hline
			QD& -617.27 & -10.65 & -625.12 & 2.80 \\
			\hline
			QW1& -481.07 & & -476.26 & 4.81 \\
			\hline
			QW2& -481.07 & & -477.34 & 3.73 \\
			\hline
		\end{tabular} 
	\end{table}
	
\noindent {\bf List of atomic configuration files}\\

\begin{enumerate}
    \item Unitcell of 1H-MoTe$_{2}$ monolayer relaxed by DF2 van der Waals functional.\\
    File name: \texttt{1\_MoTe2\_1H\_DF2\_relaxed.POSCAR}
    \item Unitcell of 1H-MoTe$_{2}$ monolayer relaxed by C09 van der Waals functional.\\	
    File name: \texttt{2\_MoTe2\_1H\_C09\_relaxed.POSCAR}
    \item Unitcell of 2H-MoTe$_{2}$ bilayer relaxed by DF2 van der Waals functional.\\
    File name: \texttt{3\_MoTe2\_2H\_BL\_DF2\_relaxed.POSCAR}
	\item Unitcell of 2H-MoTe$_{2}$ bilayer relaxed by C09 van der Waals functional.\\
	File name: \texttt{4\_MoTe2\_2H\_BL\_C09\_relaxed.POSCAR}
	\item Unitcell of 1T-MoTe$2$ monolayer relaxed by DF2 van der Waals functional.\\
	File name: \texttt{5\_MoTe2\_1T\_DF2\_relaxed.POSCAR}
	\item Unitcell of 1T-MoTe$_{2}$ monolayer relaxed by C09 van der Waals functional.\\
	File name: \texttt{6\_MoTe2\_1T\_C09\_relaxed.POSCAR}
	\item Unitcell of 1T-MoTe$_{2}$ bilayer relaxed by DF2 van der Waals functional.\\ 
	File name: \texttt{7\_MoTe2\_1T\_BL\_DF2\_relaxed.POSCAR}
	\item Unitcell of 1T-MoTe$_{2}$ bilayer relaxed by C09 van der Waals functional.\\
	File name: \texttt{8\_MoTe2\_1T\_BL\_C09-relaxed.POSCAR}
	\item Unitcell of 1T'-MoTe$_{2}$ monolayer relaxed by DF2 van der Waals functional.\\
	File name: \texttt{9\_MoTe2\_1T'\_ML\_DF2\_relaxed.POSCAR}
	\item Unitcell of 1T'-MoTe$_{2}$ monolayer relaxed by DF2 van der Waals functional.\\ 
	File name: \texttt{10\_MoTe2\_1T'\_ML\_C09\_relaxed.POSCAR}
	\item Unitcell of 1T'-MoTe$_{2}$ bilayer relaxed by DF2 van der Waals functional.\\
	File name: \texttt{11\_MoTe2\_1T'\_BL\_DF2\_relaxed.POSCAR}
	\item Unitcell of 1T'-MoTe$_{2}$ bilayer relaxed by C09 van der Waals functional.\\
	File name: \texttt{12\_MoTe2\_1T'\_BL\_C09\_relaxed.POSCAR}
	\item 1H+1T' MoTe$_{2}$, shown in Fig1b, relaxed by C09 van der Waals functional.\\
	File name: \texttt{13\_1H+1T'\_transition\_C09\_relaxed.POSCAR}
	\item C3 4-8-4 defect in MoTe$_{2}$, shown in Fig1b, relaxed by C09 van der Waals functional.\\
	File name: \texttt{14\_C3\_4-8-4\_C09\_relaxed.POSCAR}
	\item Mirror symmetric grain boundary in MoTe$_{2}$, shown in Fig2b, relaxed by C09 van der Waals functional.\\ 
	File name: \texttt{15\_Mirror\_symmetric\_GB\_defect\_C09\_relaxed.POSCAR}
	\item Quantum dot (QD) defect in MoTe$_{2}$, shown in Fig3e, relaxed by C09 van der Waals functional.\\ 
	File name: \texttt{16\_QD\_defect\_C09\_relaxed.POSCAR}
    \item Quantum wire 1 (QW1) defect in MoTe$_{2}$, shown in Fig4e, relaxed by C09 van der Waals functional.\\ 
	File name: \texttt{17\_QW1\_rot\_defect\_C09\_relaxed.POSCAR}
	\item Quantum wire 2 (QW2) defect in MoTe$_{2}$, shown in Fig4h, relaxed by C09 van der Waals functional.\\ 
	File name: \texttt{18\_QW2\_rot\_defect\_C09\_relaxed.POSCAR}
	\item Butterfly-like defect in MoTe$_{2}$, shown in Fig5c-d(ii), relaxed by C09 van der Waals functional.\\ 
	File name: \texttt{19\_Butterfly\_C09\_relaxed.POSCAR}
	\item MoTe$_{2}$ with 2 Te vacancies, shown in Fig5c(iii), relaxed by C09 van der Waals functional.\\
	File name: \texttt{20\_MoTe2\_with\_2Te\_vac.POSCAR}
\end{enumerate}

\end{document}